\documentclass[rmp,superscriptaddress,preprint]{revtex4-1}

\usepackage{amsmath, amsthm, amscd, amssymb}
\usepackage{graphicx, braket}

\begin{document}

\title{ Statistical thermodynamics for a non-commutative special relativity:
Emergence of a generalized quantum dynamics}

\author{\large \bf Kinjalk Lochan}
\email{kinjalk@tifr.res.in} \affiliation{Tata Institute of Fundamental Research\  Homi Bhabha Road\ Mumbai 400005\ India} 

\author{\large \bf Seema Satin}
\email{seema.satin@gmail.com}
\affiliation{Inst. of Math. Sciences\ IV Cross Road\ CIT Campus\ Taramani\ Chennai 600 113
\ India}

\author{\large \bf Tejinder P. Singh}
\email{tpsingh@tifr.res.in}
\affiliation{Tata Institute of Fundamental Research\ Homi Bhabha Road\  Mumbai 400005\ India}

\date{\today}

\maketitle

\centerline{\bf ABSTRACT}
\noindent There ought to exist a description of quantum field theory which does not depend on an external classical time. To achieve this goal,  in a recent paper we have proposed a non-commutative special relativity in which space-time and matter degrees of freedom are treated as classical matrices with arbitrary commutation relations, and a space-time line element is defined using a trace. In the present paper, following the theory of Trace Dynamics, we construct a statistical thermodynamics for the non-commutative special relativity, and show that one arrives at a generalized quantum dynamics in which space and time are 
non-classical and have an operator status. In a future work, we will show how standard quantum theory on a classical space-time background is recovered from here.

\tableofcontents

\newcommand{\be}{\begin{equation}}
\newcommand{\ee}{\end{equation}}
\newcommand{\F}{\mathbf{F}}
\newcommand{\y}{\hat{y}}
\newcommand{\h}{\mathbf{H}}
\newcommand{\N}{\mathbf{N}}
\newcommand{\p}{\hat{p}}

\newcommand{\x}{\hat{x}}
\newcommand{\T}{\hat{t}}
\newcommand{\G}{\mathbf{G}}

\section{Introduction}

Quantum field theory depends on an external classical space-time, in order to be able to describe evolution. We have argued elsewhere ~\cite{singh:2006, singh:2009, singh:2011} that there should exist a description of quantum theory which does not depend on an external classical time, and which should reduce to standard quantum theory as and when an external spacetime is available. In a recent paper ~\cite{Lochan-Singh:2011} we have initiated work for developing such a description, using ideas borrowed from the theory of Trace Dynamics ~\cite{Adler:04}, and by constructing what we called a non-commutative special relativity.  In the present article, we carry this work further, and obtain a generalized quantum dynamics, in which time has an operator status. In a future paper, we will demonstrate how standard quantum theory, in which time is classical, is recovered from this generalized quantum dynamics [GQD]. It should be emphasized at the outset that GQD as a theory is {\it different} from quantum dynamics: for instance, one would not use GQD to describe known laboratory experiments, nor is it clear at this stage as to what experiments could test GQD (apart from possible indirect tests which we will mention at the end of the article). Below we give a brief summary of Trace Dynamics, of non-commutative special relativity, and we give an overview of the work done in the present article.

The goal of Trace Dynamics is to arrive at quantum theory, not by quantizing a known classical theory according to the standard rules, but by {\it deriving} quantum theory from an underlying theory 
~\cite{Adler-Millard:1996}. An external classical spacetime is taken as given.

Trace Dynamics is a classical dynamics of $N\times N$ non-commuting matrices $q_r$  [equivalently operators] whose elements belong to even sector of the graded Grassmann algebra [in which case the matrices are called bosonic] or odd sector of the Grassmann algebra [fermionic matrices]. The matrices are assumed to be finite dimensional here but extension to infinite dimensions is in principle possible ~\cite{Adler:04}.
A key ingredient is the trace derivative -  the derivative with respect to an operator, of the trace of a polynomial made out of the matrices, defined by first varying, and then cyclically permuting inside the trace. The Lagrangian of the theory is the trace of a polynomial function of the matrix variables $\{q_r\}$ and their time-derivatives $\{\dot q_r\}$. Conjugate momenta $p_r$ are defined as in classical dynamics; Lagrangian dynamics is derived from an action principle, and a conserved trace Hamiltonian and Hamilton's equations are constructed. Note that although the coordinate-coordinate, coordinate-momentum and momentum-momentum commutators are in general non-zero, this is a continuum [or classical] theory, because these commutators take values which are arbitrary, and not fixed.  Nonetheless, as a result of possessing global unitary invariance, the theory admits a Poincare invariant conserved Noether charge, which is equal to the sum of the bosonic commutators $[q,p]$ minus the sum of the fermionic anticommutators $\{q,p\}$:
\begin{equation}
 \tilde{C} = \sum_{r\in B}[\hat{q}_r,\hat{p}_r] -\sum_{r\in F}\{\hat{q}_r,\hat{p}_r\} .
\end{equation}
This anti-self-adjoint matrix-valued conserved quantity, which has the dimensions of action, and which we call the Adler-Millard charge \cite{Adler-Millard:1996}, plays a profound role in the emergence of quantum theory. If $q$ and $p$ were real numbers (as in ordinary classical mechanics) instead of being matrices, this conserved quantity will be trivially zero - hence the novelty of the matrix classical dynamics.       

Since one is not interested in examining the dynamics at this level of precision, a statistical mechanics of this classical dynamics is developed, and a probability distribution derived by maximizing the entropy of the canonical ensemble, subject to the constancy of the conserved quantities. 
Assuming that the ensemble does not prefer any one state in the Hilbert space over the other, it can be shown that the canonical average of the Adler-Millard constant can be written as a real number $D$ times a diagonal, traceless, anti-self-adjoint matrix $i_{eff}$ whose elements are $\sqrt{-1}$ or $-\sqrt{-1}$. It turns out that this probability distribution is invariant under only that subset of global unitary transformations which commute with $i_{eff}$. This subset provides the largest symmetry group which allows for a non-zero value of the averaged A-M constant. Hence, in all subsequent considerations, one deals only with the projections $x_{eff}$ of dynamical variables $x$ : the $x_{eff}$ commute with $i_{eff}$. A Ward identity [analogous to the equipartition theorem] is shown to hold for any polynomial function $W,$ as a consequence of requiring that the integration phase space measure is invariant under a constant shift of the dynamical matrix variables. The identity simplifies significantly under the following two assumptions : (i) the fundamental energy scale in the problem is Planck scale, and we are for now interested only in examining the dynamics at much lower energy scales, (ii) the A-M constant appearing in the identity is replaced by its canonical average. Now, if the polynomial $W$ is chosen to be the Hamiltonian, the Ward identity implies that the canonical averages of the effective dynamical variables obey the standard Heisenberg equations of motion of quantum dynamics. If $W$ is chosen to be a dynamical variable $x$ itself, the same Ward identity implies the standard commutation/anti-commutation relations of quantum theory, {\it provided} we identify the coefficient $D$ appearing in the A-M constant with Planck's constant $\hbar$. The transition from the Heisenberg picture to the Schr\"{o}dinger picture is made as in ordinary quantum theory. Thus, with a few assumptions (some plausible, and some that will require a better understanding of Trace Dynamics) one finds quantum dynamics emerging from an underlying classical matrix theory which possesses a global unitary invariance. This happens because in the thermodynamic approximation the Adler-Millard constant has been equipartitioned over all the degrees of freedom.

Trace Dynamics is remarkable in its scope and achievement; yet it takes an external classical time as given. This should be regarded as an approximation: let us imagine a situation in which there are no external classical matter fields, and all matter is quantum. The induced spacetime metric will exhibit quantum fluctuations. Now, according to the Einstein hole argument \cite{Christian:98}, a spacetime manifold in a generally covariant theory requires a definite classical metric to exist on it, in order for us to be able to attach a physical meaning to the points of spacetime. If the metric has quantum fluctuations, the underlying spacetime manifold is destroyed, taking away from us the privilege of a pre-existing classical time. Nonetheless, one should be able to describe quantum dynamics, and hence the need for going beyond Trace Dynamics. Fortunately, the theory provides hints for its own extension. We shall assume that not only the matter degrees of freedom are to be described by matrix variables, but the same is true of spacetime points - the coordinates $\hat{x}^{\mu}=({\bf x},t)$ become matrices too. In other words, the location of every particle is described by operators $\hat{y}^{\mu}=(\hat{q}, \hat{t})$. Evolution is described with respect to a proper time $s$ (a real number) whose origin is the following.  We introduce a non-commutative Minkowski spacetime in which coordinates obey arbitrary non-commutation relations \cite{Lochan-Singh:2011}, and a line-element is defined by taking a trace:
\begin{equation}
ds^{2} = Trd\hat{s}^2\equiv Tr[d\hat{t}^2 - d\hat{x}^2 - d\hat{y}^2 - d\hat{z}^2].
\end{equation}
A novel aspect is that Poincar\'e invariance continues to hold. The interpretation is that the dynamics is invariant under a generalized set of Lorentz transformations when the spacetime coordinates do not commute with each other.  The external classical time has been removed by raising time to the status of an operator. We call this a non-commutative special relativity, and a dynamics can be constructed, following the techniques of Trace Dynamics. Bosonic and fermionic matter degrees of freedom $\hat{y}^{\mu}$ `live' on this non-commutative spacetime. In complete analogy with the construction of Adler and Millard, a Noether charge corresponding to
global unitary invariance of the Trace Hamiltonian ${\cal{H}}$ exists, for the matter degrees of freedom 
$\hat{y}^{\mu}$. It is given by
\begin{equation}
 \hat{Q} = \sum_{r\in B}[\hat{y}_r,\hat{p}_r] -\sum_{r\in F}\{\hat{y}_r,\hat{p}_r\} .
\end{equation}

Essentially one repeats the Trace Dynamics construction, with the important generalization that with every particle there is associated, besides the conjugate matrix pair $(q,p)$, also a conjugate pair $(E, \hat{t})$, where $E$ is the energy operator conjugate to $\hat{t}$. The Trace Lagrangian, action, Lagrange equations of motion, Trace Hamiltonian and Hamilton's equations of motion are constructed as before. The theory obeys global unitary invariance, and hence admits a generalized Adler-Millard conserved Noether charge which also includes commutators/anti-commutators of $E$ and $\hat{t}$. This generalized charge can be shown to be invariant under the generalized Poincar\'e transformations.

The goal and content of the present paper is the following. As in Trace Dynamics, one constructs a statistical mechanics for the non-commutative special relativity described above, and obtains an equilibrium canonical distribution by maximizing the Boltzmann entropy, subject to the constraint that certain quantities such as the Trace Hamiltonian and the generalized Adler-Millard charge are conserved. A Ward identity is derived as a result of translational invariance of the phase space measure. Approximations are made that we want to observe physics much below the fundamental (Planck) scale, and that the Adler-Millard charge be replaced by its canonical average, in the Ward identity. The emergent dynamics is a quantum-theory like description of matter and spacetime degrees of freedom. The standard commutation/anti-commutation relation between position and momenta hold, but these are now accompanied by an analogous 
commutation/anti-commutation relation between energy and operator time. In this sense, we do not have a classical spacetime background, as 
{\it all} matter and spacetime degrees of freedom have operator status. Heisenberg equations of motion hold, in which evolution is with respect to the proper time $s$, and now the Trace Hamiltonian is a function also of the operator time $\hat{t}$ associated with each particle.
There is an equivalent generalised Schr\"{o}dinger picture, in which the wave-function evolves with respect to the proper time, and is a function of operator time variables as well. We can think of this theory as a Generalised Quantum Dynamics, a generalisation of quantum theory, in the absence of classical time. In a future work, we will demonstrate the emergence of a classical space-time from here, and explain how the energy-time commutator vanishes in that limit.

Because non-commutative special relativity possesses a Poincar\'e invariance and a conserved 
Adler-Millard charge, the construction of its statistical thermodynamics very closely follows the development of the corresponding steps in Trace Dynamics. The description in the following sections is patterned after the book of ~\cite{Adler:04} and the interested reader will find additional details there: the procedure described there for configuration variables is repeated here, except that the variables now include a time operator associated with every degree of freedom. It is interesting that the inclusion of this additional freedom does not have an impact on the formal construction of the statistical thermodynamics.

\section{The Phase space}
We began by defining `space-time' in operator form by matrix degrees of freedom $\y^{\mu}(s) \equiv 
\{\T(s), \x^i(s)\}$ as in ~\cite{Lochan-Singh:2011}, where $s$ denotes the proper-time  and is used as a parameter. 
The motion of `test particles' on this `space-time'  (which can be thought to be parameterized by one-parameter family of trajectories),
can be described by a trace Lagrangian defined as 
\be
\mathcal{L}  = \sqrt{Tr \eta_{\mu \nu} d \y^\mu d \y^\nu}
\ee
 The momenta conjugate to $\y_r^{\mu}$ are given by 
\be
\p_\mu = \frac{\delta L}{\delta \dot{\y}^\mu}
\ee 
where a `dot' denotes derivative with respect to proper time $s$.
We denote $(\y_r^\nu,\p_{r\nu }) \equiv (z_r(\nu), z_{r+1}(\nu)) $ which describe the phase space variables,
while $r$ marks the one parameter family of systems moving on geodesics
covering the underlying `manifold'. These represent and describe the matter degrees of freedom and could be either bosonic or fermionic matrices.
We next prove that in this matrix model `space-time' dynamics there exists an invariant phase space
volume element. Thus this gives us an  analogue of Liouville theorem at the trace-dynamics level.
\subsection{The Liouville Theorem}
Let  $(z_r)_{mn}$ denote the general matrix element of the operator $z_r$, which 
can be decomposed  into its real and imaginary parts,
\[(z_r)_{mn} = (z_r)^0_{mn} + i (z_r)^1_{mn}. \] 
Define the phase space measure as follows
\begin{eqnarray}
d \mu & =& \Pi_R d\mu^R \\
d \mu^R & = & \Pi_{r,m,n} d(z_r)_{mn}^R 
\end{eqnarray}
where $R = 0,1$ denote the real and imaginary parts.

Now we show that the phase space volume element is invariant under general canonical
transformations given by
\begin{eqnarray}
\delta \p_{r \mu} & = & - \frac{\delta \G}{\delta \y_r^{\mu}} \\
\delta \y_r^{\mu} & = & \epsilon_r \frac{\delta \G}{\delta \p_{r \mu}},
\end{eqnarray}
where ${\bf G}$, the trace of a polynomial $G$ made out of the matrices,  is the generator of canonical transformation and $\epsilon_r = \pm 1$ [+1 for bosonic matrices, and -1 for fermionic matrices].
The above equations can be written in compact form as 
\be 
\delta z_r = \sum_s \omega_{rs} \frac{\delta \G}{\delta z_s}
\ee
where $\omega_{rs} = diag(\Omega_B,\Omega_B,\Omega_B,..., \Omega_F,.., \Omega_F ),$ with
$$\Omega_B=\left(
\begin{array}{ll}
0   & -1\\
1 &  \ \ 0 
\end{array}
\right),$$
and
$$\Omega_F=-\left(
\begin{array}{ll}
0 & 1\\
1 & 0 
\end{array}
\right).$$
If we denote any arbitrary trace functional  by $\F$, then, under canonical 
transformation upto first order we obtain, in terms of the generalized Poisson bracket $\{... , ...\}_{GPB}$,
\be
\delta \F= \{\F,\G\}_{GPB}.
\ee
Next we note that the matrix variables $\tilde{z}_r = z_r + \delta z_r$ under general canonical transformation are related 
to the original $z_r$ by
\be \label{eq:cantrans}
(\tilde{z_r})^R_{mn} = (z_r)^R_{mn} + \sum_s \omega_{rs} \left(\frac{\delta \G}{
\delta z_s}\right)^R_{mn}.
\ee
From the definition of trace derivative ~\cite{Adler:04}
\be
\delta \G = Tr \sum_s \frac{\delta \G}{\delta z_s} \delta z_s
\ee
and also using the reality of $\G$ one can write
\be
\delta \G = \sum_{s,m,n,R} \epsilon^R\left(\frac{\delta \G}{\delta z_s}\right)^R_{mn}
(\delta z_s)^R_{nm}
\ee
where $\epsilon^0 = 1 $ and $\epsilon^1 = -1$.
 Hence
\be
\left(\frac{\delta \G}{\delta z_s}\right)^R_{mn} = \epsilon^R \frac{\partial \G}
{\partial (z_s)^R_{nm}}.
\ee
Equation (\ref{eq:cantrans}) can be written as
\be
(\tilde{z_r})^R_{mn} = (z_r)^R_{mn} + \sum_s \omega_{rs} \epsilon^R \frac{
\partial \G}{\partial (z_s)_{nm}^R}.
\ee
Differentiating the above w.r.t $(z_{r'})^R_{m'n'}$ we get
\[ \frac{\partial (\tilde{z_r})^R_{mn}}{\partial (z_{r'})^R_{m'n'}}
= \delta_{rr'}\delta_{mm'} \delta_{nn'} + \sum_s \omega_{rs} \epsilon^R
\frac{\partial^2 \G}{\partial (z_s)^R_{nm} \partial ( z_{r'})^R_{m'n'}} .\]
The Jacobian of the matrix is given by 
\begin{eqnarray}
 J & = & 1 + \Delta  \\
\Delta & = & \sum_{r,s,m,n} \omega_{rs} \epsilon^R \frac{ \partial^2 \G}
{\partial (z_s)^R_{nm} \partial (z_r)^R_{mn}}. 
\end{eqnarray}
Interchanging summation indices $r$ and $s$, and also $m$ and $n$,
\be
\Delta = \sum_{r,s,m,n} \omega_{sr} \epsilon^R \frac{ \partial^2 \G}{
\partial (z_r)^R_{mn} \partial (z_s)^R_{nm}}. \label{J}
\ee
For bosonic $r$ and $s$ it is known that ~\cite{Adler:04} $\omega_{sr} = - \omega_{rs}$.
Hence
\be
\frac{ \partial^2 \G }{\partial (z_r)^R_{mn} \partial (z_s)^R_{nm}}
= \frac{\partial^2 \G}{\partial(z_s)^R_{nm} \partial (z_r)^R_{mn}}.  \label{J1}
\ee
For fermionic $r$, $s$, $\omega_{sr} = \omega_{rs} $,
\be
\frac{ \partial^2 \G}{\partial (z_r)^R_{mn} \partial(z_s)^R_{nm}} =
- \frac{ \partial^2 \G}{\partial (z_s)^R_{nm} \partial (z_r)^R_{mn}}. \label{J2}
\ee 
Thus we see that $\Delta = - \Delta$; hence $\Delta$ vanishes and the Jacobian
is unity. Thus we see that the matrix operator phase space
integration measure in the above case is invariant under general canonical
transformations. However, we argued in ~\cite{Lochan-Singh:2011} that for the `non-commutative special relativistic' theory
all $\hat{y}^{\mu}$s are either self-adjoint or anti-self-adjoint, but not mixed. So, we need to show that there exists an invariant measure on
the phase space when the phase space variables are constrained to be self-adjoint (or anti-self-adjoint). 
This is not difficult to show if one follows the scheme built by Adler ~\cite{Adler:04}.

Using (\ref{J1}) and (\ref{J2}) one sees that diagonal and off-diagonal elements in the sum (\ref{J}) vanish separately. Therefore, adjointness 
restriction can be handled effectively. For our case, (\ref{J}) becomes
\be
\Delta = \sum_{r,s,m} \omega_{sr} \frac{ \partial^2 \G}{\partial (z_r)^{0/1}_{mm} \partial (z_s)^{0/1}_{mm}} + 
\sum_{r,s,m<n,R} \omega_{sr} \epsilon^R \frac{ \partial^2 \G}{
\partial (z_r)^R_{mn} \partial (z_s)^R_{nm}}, \label{JwithA}
\ee
with $0/1$ marking for the self-adjoint/anti-self-adjoint case. Then as argued before, symmetry properties of $\omega_{sr}$ make $\Delta$ vanish, leaving 
us with the invariant measure on the adjoint-restricted phase space.

\section{Canonical Ensemble}
We do not require the specific form of the measure, but only the Liouville theorem
 and the property that the measure is invariant under infinitesimal matrix 
operator shifts $\delta z_r $,
\[ d \mu[{z_r +\delta z_r}] = d \mu[{z_r}] .\]
We denote the equilibrium phase space density distribution by $\rho[{z_r}] 
\geq 0 $.
Then, the infinitesimal probability density of finding the system in operator phase
space volume element $d \mu$ is given by
\[  dP = d \mu[{z_r}] \ \rho[{z_r}]. \]
Also,
\[ \int dP = 1 = \int d \mu[{z_r}] \rho[{z_r}]. \]
Equilibrium implies that $\dot{\rho} = 0$. Therefore, the phase space distribution function can be built with the
conserved charges of the theory apart from some constant coefficients. In case of Poincar\'{e}  invariant theories there will also be
a conserved trace energy momentum density ${\cal T}^{\mu \nu}$. The integration of ${\cal T}^{\mu \nu}$ with a further restricted phase space measure
(constrained along $\x^i$ plane) gives us additional conserved charges ${\bf \vec{K},\vec{J}, \vec{P}}$ corresponding to `boosts', 
`rotations' and `translations'. However, the Poincar\'{e} invariant nature of the theory helps us in suitably working in a frame where these quantities might
be taken to vanish. 
Thus the general equilibrium 
distribution takes the form for a Poincar\'{e} invariant theory,
\[ \rho\equiv\rho(\hat{Q}, \mathbf{H}, \mathbf{N}) \]
where ${\bf H}$ is the conserved Trace Hamiltonian and ${\bf N}$ is the conserved fermion number.
We can include  a traceless anti-self-adjoint operator parameter $\tilde{\lambda} $ as the Lagrange multiplier corresponding to conserved 
operator $\hat{Q}$ ~\cite{Lochan-Singh:2011} 
and real number parameters $ T $ and $\eta$
which correspond to conserved quantities $\mathbf{H} $ and $\mathbf{N}$ and we can write
\be
\rho = \rho(\hat{Q}, \tilde{\lambda}; \mathbf{H}, T; \mathbf{N}, \eta ).
\ee
Since we have suitably defined a measure and a distribution function on the phase space we define
the {\it ensemble average} for a general operator $B$ 
\[ \langle B \rangle _{AV} = \int d \mu \ \rho B. \]
We will keep in our mind that unlike the classical ensemble average this {\it ensemble average}  is a map 
\[\langle.\rangle_{AV} :{\cal G}_M \rightarrow {\cal G}_M .\]
Next we try to obtain the functional form of $\rho$  in the canonical ensemble as
follows:
\be
 S = - \int d \mu \rho \log \rho
\ee 
is defined as the entropy of the system. This entropy is extremized subject to the following conservation constraints
\begin{eqnarray}
\int d \mu \rho & = & 1 \\
\int d \mu \rho \hat{Q} & = & \langle\hat{Q}\rangle_{AV} \\
\int d \mu \rho \mathbf{H} & = & \langle \mathbf{H} \rangle_{AV} \\
\int d \mu \rho \mathbf{N} & = & \langle\mathbf{N} \rangle_{AV}. 
\end{eqnarray}
Imposing constraints with Lagrange multipliers $\theta, \tilde{\lambda}, T, 
\eta $ we extremize the following expression
\be
\mathcal{F} = \int d \mu \log \rho + \theta \int  d \mu \rho + \int d \mu \rho
Tr \tilde{\lambda} \hat{Q} + T \int d \mu \rho \mathbf{H} + \eta \int d \mu \rho \mathbf{N}
\ee
Taking derivative of the above with respect to $\rho$ and equating it to zero
gives:
\be
\rho = \exp(-1-\theta - Tr \tilde{\lambda} \hat{Q} - T \h - \eta \N)
\ee
Putting in the condition that $\rho$ should be normalized to unity, gives
\begin{eqnarray} \label{eq:partition}
\rho & =  & Z^{-1} \exp(-Tr \tilde{\lambda} \hat{Q} - T \h - \eta \N) \\
Z & = & \int d \mu \exp(- Tr \tilde{\lambda} \hat{Q} - T \h - \eta \N), 
\end{eqnarray}
where $Z$ is the partition function of the system.
Thus we see that
\be 
S = -\langle\log \rho\rangle_{AV} = \log Z + Tr \tilde{\lambda} \langle\hat{Q}\rangle_{AV} + 
T \langle\h\rangle_{AV}
 + \eta \langle\N\rangle_{AV}. 
\ee
The above equation  can be re-written as 
\be \label{eq:entropy}
S = \log Z - Tr \tilde{\lambda} \frac{ \delta \log Z}{\delta \tilde{\lambda}}
- T \frac{\partial \log Z}{\partial T} - \eta \frac{\partial \log Z}{\partial
\eta}, \ee
since the ensemble averages are given by
\begin{eqnarray}
\langle\hat{Q}\rangle_{AV} & = &  - \frac{ \delta \log Z}{\delta \tilde{\lambda}} \\ \nonumber
\langle\h\rangle_{AV} & = & - \frac{\partial \log Z}{\partial T } \\ \nonumber
\langle\N\rangle_{AV} & = & - \frac{ \partial \log Z}{\partial \eta}.
\end{eqnarray}
From equation (\ref{eq:entropy}) it is clear that the entropy can be determined solely by the partition function. Therefore we can have a 
thermodynamic description of `non-commutative special relativity' in terms of the partition function. Also, the variances in the meaningful observable quantities
can be obtained from the partition function in the usual fashion.

For an operator ${\cal O}$ made from phase-space variables using only $c$-number coefficients, it can be shown that its ensemble average takes the form
\begin{equation}
 \langle {\cal O} \rangle_{AV}=\int d\mu \rho {\cal O}=F_{\cal O}(\tilde{\lambda}). \label{AvgProp}
\end{equation}
This can be verified by first doing a unitary transformation over (\ref{AvgProp}) and then using invariance of integration measure.
An obvious outcome of this is that any geometric operator made up of space-time phase space variable commutes with $\tilde{\lambda}$. Owing to
the fact that an anti-self-adjoint $\tilde{\lambda}$ can be diagonalized by a unitary transformation, and using (\ref{AvgProp}), for 
${\cal O}=\hat{Q}$,
\begin{equation}
 \langle\hat{Q}\rangle_{AV} = i_{eff}D_{eff}, \label{Qav1}
\end{equation}
with $i_{eff}$ some unitary diagonal phase operator and $D_{eff}$ some real, positive, diagonal magnitude operator. Now, this `polar form' of
$  \langle\hat{Q}\rangle_{AV}$ together with conditions of being traceless, anti-self-adjoint and symmetric in all eigen-directions ~\cite{Adler:04} implies for the Adler-Millard charge that 
\begin{eqnarray}
 \langle\hat{Q}\rangle_{AV} = i_{eff} {\hslash}, \label{Qav2}\\
 Tr i_{eff} =0,
\end{eqnarray}
 with a form for $i_{eff} = diag (i,-i,i,-i,...i,-i), $ and a constant ${\hslash}$ with the dimensions of action.

Now, using (\ref{Qav1}), (\ref{Qav2}) and the fact that $ \langle\hat{Q}\rangle_{AV}$ commutes with $\tilde{\lambda}$, gives
\begin{equation}
 \tilde{\lambda}=\lambda i_{eff},
\end{equation}
with a $c-$number $\lambda$. However, presence of $Tr \tilde{\lambda} \hat{Q}$ term in the partition function breaks the invariance of the
underlying dynamics under global unitary transformation to a residual subset $U_{eff}$ of global unitary transformations, which commutes with $i_{eff}$. Thus, 
an important implication of the canonical ensemble is reduction of the symmetry group. However, the cost of introducing a non-vanishing $\langle\hat{Q}\rangle_{AV}$ is not enormous,
in the sense that we are still inside the subset of unitary transformation {\it (consistent with the Heisenberg dynamics we want to achieve)}. From now
on, we will be left with this residual unitary invariance of underlying `space-time' degrees of freedom. The residual unitary invariant average of a 
polynomial $R_{eff}$ commuting with $i_{eff}$, which is made up of $i_{eff}$ and phase-space variables is defined as
\begin{equation}
 \langle R_{eff}\rangle_{AV}=\frac{\int d[U_{eff}]d\hat{\mu} \rho R_{eff}}{\int d[U_{eff}]d\hat{\mu} \rho}, \label{groupavg}
\end{equation}
where $d\hat{\mu}$ is integration measure on phase space with a fixed global unitary transformation $U_{eff}$, and $d[U_{eff}]$ being Haar measure on the
residual subgroup.
\subsection{Breaking of residual unitary transformation invariance}
As already argued in ~\cite{Adler:04}, if one has unrestricted average of 
$R_{eff}$ over the canonical ensemble, the resulting  $\langle R_{eff}\rangle_{AV} $ will lack
the interesting matrix operator structure. We would like (\ref{groupavg}) to maintain this structure. For that to happen we would like to restrict the integration 
over the group so as to keep an overall global unitary transformation $U_{eff}$ fixed. As we will see soon, Poincar\'{e} transformations of the underlying
space-time dynamics which are also canonical transformations, will manifest themselves as global unitary transformation $U_{eff}$ in the emergent canonical 
ensemble. We need to break the residual unitary invariance otherwise symmetry group  will also be averaged over in integration.
Therefore, unrestricted integration over $U_{eff}$ is avoided.

 In order to do so, if trace dynamics is assumed irreducible, fixing global unitary transformation on one canonical pair $r=R,R+1$ does the job. So,
the restricted measure on phase space from now on will exclude one pair. Although unitary fixing keeps $i_{eff}$ diagonal (as unitary fixing is done
within the subgroup of unitary transformations commuting with it), and restricted canonical averages
of operators still commute with $\tilde{\lambda}$, breaking the residual unitary invariance results in fluctuation in $\langle\hat{Q}\rangle_{\hat{AV}}$ about its
 unrestricted canonical average values. For unitary fixed
$$[\tilde{\lambda}, \langle\hat{Q}\rangle_{\hat{AV}}]=0,$$
the general form for $\langle\hat{Q}\rangle_{\hat{AV}}$ will be
\begin{equation}
 \langle\hat{Q}\rangle_{\hat{AV}}=i_{eff}\hslash+\frac{1}{2}(\tau_0+\tau_3)\Delta_+ + \frac{1}{2}(\tau_0-\tau_3)\Delta_-,
\end{equation}
with $\tau_i$ being Pauli matrices whereas $\Delta_{\pm}$ are anti-self-adjoint traceless matrices which are non-zero in general. Effects of these correction
terms will be studied in detail elsewhere.

\section{A Ward Identity and Emergence of Generalized Quantum Dynamics}
With a restricted integration measure, the averages are just rewritten as
\begin{equation}
 \langle {\cal O} \rangle_{\hat{AV}}=\int d\hat{\mu} \rho_j {\cal O}, \label{AvgProp1}
\end{equation}
with inclusion of a source term $Trj_rz_r$ in the distribution function
\begin{eqnarray} \label{eq:partition1}
\rho_j & =  & Z_j^{-1} \exp(-Tr \tilde{\lambda} \hat{Q} - T \h - \eta \N-\sum_r Tr j_rz_r) \\
Z_j & = & \int d\hat{\mu} \exp(- Tr \tilde{\lambda} \hat{Q} - T \h - \eta \N -\sum_r Tr j_rz_r), 
\end{eqnarray}
that tells about the average properties of `space-time' degrees of freedom. Now, the restricted integration measure on phase space is invariant under 
constant shift : $z_r\rightarrow z_r+\delta z_r$ of any dynamic variable apart from the unitary fixed one. Making use of this invariance we can write
for a bosonic polynomial $W$ made from space-time variables, under shift of one variable $z_s$
\begin{eqnarray}
 \delta_{z_s}Z_j\langle Tr\{\hat{Q},i_{eff}\}W\rangle _{\hat{AV},j}=0=\int d\hat{\mu} \delta_{z_s}[\exp(- Tr \tilde{\lambda} \hat{Q} - \nonumber\\
T \h - \eta \N -\sum_r Tr j_rz_r)Tr\{\hat{Q},i_{eff}\}W]. \label{PreWardI} 
\end{eqnarray}
For a polynomial $W$ made up of space-time variables,
\begin{equation}
\delta_{z_s} W =\sum_l W_s^{Ll}\delta z_s W_s^{Rl},
\end{equation}
where superscript $l$ marks a monomial in the polynomial and $R,L$ mark right and left decomposition of the monomial under infinitesimal variation in $z_s.$
Therefore,
\begin{equation}
\frac{\delta {\bf W}}{\delta z_s}=\sum_l\epsilon_l W_s^{Rl}W_s^{Ll}. \label {Wvar}
\end{equation}

Now, (\ref{PreWardI}) again on following Adler's procedure ~\cite{Adler:04}, leads us to a Ward identity for the variable $z_{r eff}$ (part of $z_r$ which commutes with $i_{eff}$),
\begin{eqnarray}
 \langle \Lambda_{r eff}\rangle_{\hat{AV}, j} &=& 0;\nonumber\\
 \Lambda_{r eff} &=& (-T\dot{z}_{r eff}+i\eta\xi_rz_{r eff}-\sum_s \omega_{rs}j_{s eff})Tr\hat{Q}i_{eff}W_{eff}\nonumber\\
                 &+& [i_{eff}W_{eff}, z_{r eff}]+\sum_{s,l}\omega_{rs}\epsilon_l(W_s^{Rl}\frac{1}{2}\{\hat{Q},i_{eff}\}W_s^{Ll})_{eff},\nonumber\\
\end{eqnarray}
with $\xi_r=0$ for bosonic conjugate pair while $\xi_r=\pm 1$ for fermionic variable/conjugate momenta and [ $]_{eff}$ of any operator gives that part
of the operator which commutes with $i_{eff}$. Also, for matrix operator $M$,
\begin{equation}
 2i_{eff}M_{eff}=\{i_{eff},M\}.
\end{equation}
 Now, for writing in a more compact form we define
\begin{eqnarray}
 {\cal D}z_{r eff} &=& (-T\dot{z}_{r eff}+i\eta\xi_rz_{r eff})Tr\hat{Q}i_{eff}W_{eff}\nonumber\\
                   &+& [i_{eff}W_{eff}, z_{r eff}]+\sum_{s,l}\omega_{rs}\epsilon_l(W_s^{Rl}\frac{1}{2}\{\hat{Q},i_{eff}\}W_s^{Ll})_{eff}, \label {Dz}
\end{eqnarray}
such that, we can write the Ward identity obtained above as,
\begin{eqnarray}
  \langle {\cal D}z_{r eff}\rangle_{\hat{AV}, j}-\sum_s \omega_{rs}j_{s eff}\langle Tr\hat{Q}i_{eff}W_{eff}\rangle_{\hat{AV}, j}=0. \label{Ward}
\end{eqnarray}
Since we have action of ${\cal D}$ on $z_{reff}$, in principle we can extend its action, using Leibniz rule, to polynomials made up of $z_{reff}$, such that
at zero sources
\begin{equation}
 \langle S_L(z_{seff})({\cal D}S(z_{reff}))S_R(z_{seff})\rangle_{\hat{AV}, 0}=0, \label{Polynomial}
\end{equation}
where the ${\cal D}$ operator splits the polynomial $S$ in left and right components, on action. With this we can extend the Ward-identity to polynomials
$S(z_{reff})$ as well. Next, we will take up different $W$s to use the identity effectively in the space-time context.
We first restrict ourselves to the following scenario:\\
{\bf (A)} The term 
$$ -T\dot{z}_{r eff} \{\hat{Q},i_{eff}\}W_{eff},$$
can be neglected on two accounts. First $z_{r eff}$ has a slow and a fast varying part. The parameter $T$ is identified with the fast scale
which is assumed to be the Planck scale. So the slow part does not change or contribute significantly over the Planck scale. Moreover, the fast varying 
part is assumed to have disjoint support from $ \{\hat{Q},i_{eff}\}=-\frac{1}{2}i_{eff}\hat{Q}_{eff}$.\\
{\bf (B)} The chemical potential $\eta$ is not a relevant parameter and can be neglected. The term 
$$i\eta\xi_rz_{r eff}Tr\hat{Q}i_{eff}W,$$
contributes only in the fermionic sector. We are restricted to an ensemble with zero ensemble average of trace fermion number.\\
{\bf (C)} We  replace $\hat{Q}_{eff}$ in $ \{\hat{Q},i_{eff}\}=-\frac{1}{2}i_{eff}\hat{Q}_{eff}$ by its 
zero source canonical average $i_{eff}\hslash$.\\
Therefore, with these assumptions and (\ref{Wvar}), Equation (\ref{Dz}) can be written neatly as,
\begin{equation}
 {\cal D}z_{r eff} =i_{eff}[W_{eff}, z_{r eff}]-\hslash\sum_s\omega_{rs}\left(\frac{\delta {\bf W}}{\delta z_s}\right)_{eff}. \label{ModWard}
\end{equation}
(i) Now, if we take $W$ as the operator Hamiltonian $H$, we are led to
\begin{equation}
 {\cal D}z_{r eff} =i_{eff}[H_{eff}, z_{r eff}]-\hslash\dot{z}_{r eff}, \label{ModWard1}
\end{equation}
the Heisenberg evolution of space-time phase space variables. We can also see from here that using Ward identity (\ref{Ward}) and 
(\ref{Polynomial}), for a polynomial $P_{eff}$
\begin{eqnarray}
\langle S_L(z_{s eff})\dot{P}_{eff} S_R(z_{s eff}) \rangle_{\hat{AV},0}=\nonumber\\
\langle S_L(z_{s eff})i_{eff}\hslash^{-1} [H_{eff}, P_{eff}]S_R(z_{s eff}) \rangle_{\hat{AV},0}
\end{eqnarray}
(ii) For $W=\sigma_vz_v$,
\begin{equation}
 i_{eff}{\cal D}z_{r eff} =[z_{r eff}, \sigma_v z_{v eff}]-i_{eff}\hslash\omega_{rv}\sigma_v. \label{ModWard2}
\end{equation}
Using (\ref{Ward}), (\ref{ModWard2}) gives the emergent canonical commutation rules for the bosonic and fermionic degrees of freedom, sandwiched between
polynomials $S_{L/R}$ and averaged over zero source ensembles. Thus we obtain, what we call effective canonical commutators of the matter degrees of freedom.
For bosonic pair
\begin{equation}
 [y^{\mu},y'^{\nu}]=0; \quad  [y^{\mu},p_{\nu}]=i_{eff}\hslash\delta^{\mu}_{\nu},
\end{equation}
while for fermionic pair
\begin{equation}
 \{y^{\mu},y'^{\nu}\}=0; \quad \{y^{\mu},p_{\nu}\}=i_{eff}\hslash\delta^{\mu}_{\nu}.
\end{equation}
This leads to non-commutativity among configuration variables and corresponding momenta of the matter degrees of freedom, at emergent canonical level. Evidently, there is included now a 
operator time - energy commutation relation. In anticipation of the standard quantum theory that will eventually emerge from here, we may identify the constant $\hbar$ with Planck's constant.\\
(iii) For ${\bf W=G}$, a generator of canonical transformation, we obtain unitary evolution of the matter degrees of freedom of non-commutative special relativity at the effective canonical averaged
level,
\begin{eqnarray}
 \langle S_L(z_{s eff})\delta z_{r eff}S_R(z_{s eff})\rangle_{\hat{AV},0}= \nonumber\\
 \langle S_L(z_{s eff})i_{eff}\hslash[G_{eff},z_{r eff}]S_R(z_{s eff}) \rangle_{\hat{AV},0} .
\end{eqnarray}

Thus, we obtain basic ingredients of a generalized quantum dynamics at the emergent level, which are \\
(i) Heisenberg equation of motion,\\
(ii) commutator/anti-commutator properties,\\
(iii) Unitary generation of canonical transformations.

We have a system whose dynamical degrees of freedom can be studied with the evolution scheme illustrated above. For that, at the emergent
level, we identify the operator polynomial $P(z_{eff})$ averaged between $S_{L/R}$ in matrix model as operator polynomial $P({\cal X}_{eff})$ at the
emergent level. 

Further, since at the fundamental matrix level, the theory is Lorentz invariant as shown in 
~\cite{Lochan-Singh:2011}, if we add another assumption of
 boundedness of $H_{eff}$ and existence of zero eigenvalue of $\vec{P}_{eff}$ corresponding to a unique eigenstate $\psi_0$, there is a 
proposed correspondence between canonical ensemble average quantities and Lorentz-invariant Wightmann functions in the emergent field theory,
$$ \psi_0^{\dagger}\langle P({z_{eff}})\rangle_{\hat{AV}}\psi_0=\langle vac |P({{\cal X}_{eff}})|vac \rangle. $$
As a result we obtain operator statements at the effective level corresponding to the three ingredients of a generalized field theory at the trace-dynamics level listed above.\\
(i) $[{\cal X}_{reff},\sigma_s {\cal X}_{seff}] = i_{eff}\hslash\omega_{rs}\sigma_s,$\\
(ii) $\dot{P} ({\cal X}_{reff})=i_{eff}\hslash^{-1}[H_{eff},P({\cal X}_{reff})],$\\
(iii) $\delta{P} ({\cal X}_{reff})=i_{eff}\hslash^{-1}[G_{eff},P({\cal X}_{reff})].$ 
\newline
We can now obtain an equivalent Schr\"{o}dinger picture corresponding to the emergent Heisenberg picture of space-time dynamics. For that, we 
define
$$U_{eff}(s)=\exp{(-i_{eff}\hslash^{-1} s H_{eff})},$$ 
such that
$$\frac{d}{ds}U_{eff}(s)=-i_{eff}\hslash^{-1}H_{eff}U_{eff}(s).$$
Then, for a Heisenberg state vector $\psi$ we form Schr\"{o}dinger picture state vector $\psi_{schr}(s)$, for space-time degree of freedom
$$ \psi_{schr}(s)=U_{eff}(s)\psi,$$
$$i_{eff}\hslash \frac{d}{ds}\psi_{schr}(s)=H_{eff}\psi_{schr}(s).$$
Thus we obtain Schr\"{o}dinger evolution for the phase-space variables at the canonical ensemble average level. However, as discussed above,
breaking of the residual unitary transformation group results in fluctuations about the canonical averages; this Schr\"{o}dinger evolution is hence a leading order description of the evolution.

\section{Discussion}

We have applied the principles of Trace Dynamics to derive a generalized quantum dynamics, which does not depend on classical time, starting from an underlying non-commutative special relativity in which space-time coordinates have an operator status. In a follow-up work, we will analyze how standard quantum theory with a classical time emerges from this generalized quantum dynamics. This will be done by continuing to follow the path of Trace Dynamics where classical dynamics is derived from quantum theory by considering fluctuations around thermodynamic equilibrium.

Recall that in Trace Dynamics quantum theory is derived from the underlying classical theory of matrices, in the thermodynamic limit.  Where there is statistical thermodynamics, there are fluctuations. The next step in the scheme is to consider statistical fluctuations around the thermodynamic approximation made above. This involves giving up the assumption that in the Ward identity the Adler-Millard constant be replaced by its average; instead the constant is replaced by its average plus a term which represents Brownian motion type fluctuations. Instead of the standard linear Schr\"{o}dinger equation of quantum theory, one now gets a stochastic non-linear Schr\"{o}dinger equation; the stochastic part is contained in the nonlinear terms which have resulted on account of including the Brownian fluctuations. Two significant results emerge. If one sets up a quantum system in an initial state which is a superposition of the eigenstates of an observable $A$ which commutes with the Hamiltonian, the stochastic average of the variance of $A$ goes to zero as the time elapsed goes to infinity. This means that the system has been driven to a definite eigenstate of $A$, though we cannot predict which one. But we can calculate the probability of every outcome. This is possible because the stochastic average of the expectation value of the projectors $\pi=|a\rangle\langle a|$ of $A$ is a constant in time. This quantity is initially equal to 
$|\langle\psi|a\rangle |^{2}$, the square of the amplitude for the system to be initially in the state $|a\rangle$. The final value of this stochastic average, as $t\rightarrow \infty$, is the probability $P_a$ for the system to be driven to the state $|a\rangle$. Hence $P_a = |\langle\psi|a\rangle |^{2}$, which is the Born probability rule, now derived from first principles. The time-scale over which the system is driven to one of the eigenstates can be calculated, and depends on the number of degrees in the system - it is astronomically large for microscopic systems, and unmeasurably small for macroscopic ones. In other words, the nonlinear stochastic fluctuations are significant for macroscopic systems, but utterly negligible for microscopic ones. This is how we are to understand the dynamical collapse of the wave-function during a quantum measurement, and the absence of position superpositions in large systems.   

Like in Trace Dynamics, in the derivation of the generalized quantum dynamics from non-commutative special relativity,  one next allows for inclusion of stochastic fluctuations of the Adler-Millard charge, in the Ward identity. This leads once again to a stochastic non-linear Schr\"{o}dinger equation, but now with some additional consequences. Consider a situation where matter starts forming macroscopic clumps [for instance in the very early universe, soon after a Big Bang]. The stochastic fluctuations become more and more important as the number of degrees of freedom in the clumped system is increased. These fluctuations force macroscopic objects to be localized, but now not only in space, but also in time! This means that the time operator associated with every particle has become classical (a 
c-number times a unit matrix). The localization of macroscopic objects comes hand in hand with the emergence of a classical spacetime, in accordance with the Einstein hole argument and the general theory of relativity. If, and only if, the Universe is dominated by macroscopic objects, as today's Universe is, can one also talk of the existence of a classical spacetime. When this happens, the proper time $s$ may be identified with classical time. The Born probability rule is still at work, randomly selecting one possible macrostate out of the many that are possible. Once the Universe has reached this classical state, it sustains itself therein, by virtue of the continual action of stochastic fluctuations on macroscopic objects, thereby permitting also the existence of a classical spacetime geometry. 

That done, we realize that although almost all of the matter in the universe is clumped into macroscopic objects, innumerable laboratory experiments [as also astronomical observations such as stellar spectra and the black-body nature of the Cosmic Microwave Background Radiation] provide inescapable evidence for the sub-dominant microscopic world where dynamics seems to have different rules. To arrive at these, we first develop and derive the concept of a classical time, as described above. Taking this classical time as {\it given}, one constructs the Trace Dynamics of the matter degrees of freedom, thus obtaining standard quantum theory in the thermodynamic approximation, and the Born probability rule and an explanation for the measurement problem by considering Brownian fluctuations around this thermodynamic approximation.

The ideas described in these few paragraphs above will be put on a firm mathematical footing in a forthcoming article.

Trace Dynamics predicts a non-linear stochastic modification of the Schr\"{o}dinger equation, with the corrections becoming more and more important as the number of degrees of freedom in the system is increased. This prediction can in principle be tested by experiments such as molecular interferometry which aim to verify the principle of linear superposition for macroscopic systems ~\cite{Adler:09}. Stochastic modifications of the generalized quantum dynamics proposed here would also lead a similar prediction.

\bigskip

\noindent {\bf Acknowledgement}: This publication was made possible through the support of a grant
from the John Templeton Foundation. The opinions expressed in this publication are
those of the authors and do not necessarily reflect the views of the John Templeton
Foundation.

\bibliography{biblioqmts2}
\end{document}